\documentclass[aip,jap,reprint]{revtex4-1}
\usepackage{graphicx}
\usepackage{ulem}
\usepackage{amssymb}
\usepackage{amsmath}
\usepackage{wasysym}
\usepackage{relsize}
\begin{document}


\newcommand{\be}{\begin{equation}}
\newcommand{\ee}{\end{equation}}
\newcommand{\bea}{\begin{eqnarray}}
\newcommand{\eea}{\end{eqnarray}}
\newcommand{\Tbar}{{\bar{T}}}
\newcommand{\En}{{\cal E}}
\newcommand{\K}{{\cal K}}
\newcommand{\U}{{\cal U}}
\newcommand{\GC}{{\cal \tt G}}
\newcommand{\Lop}{{\cal L}}
\newcommand{\DB}[1]{\marginpar{\footnotesize DB: #1}}
\newcommand{\q}{\vec{q}}
\newcommand{\kt}{\tilde{k}}
\newcommand{\Lopn}{\tilde{\Lop}}
\newcommand{\noi}{\noindent}
\newcommand{\ovn}{\bar{n}}
\newcommand{\ovx}{\bar{x}}
\newcommand{\ovE}{\bar{E}}
\newcommand{\ovV}{\bar{V}}
\newcommand{\ovU}{\bar{U}}
\newcommand{\ovJ}{\bar{J}}
\newcommand{\calE}{{\cal E}}
\newcommand{\ovphi}{\bar{\phi}}
\newcommand{\zt}{\tilde{z}}
\newcommand{\ttl}{\tilde{\theta}}
\newcommand{\nuv}{\rm v}
\newcommand{\ds}{\Delta s}
\newcommand{\fn}{{\small {\rm  FN}}}
\newcommand{\cc}{{\cal C}}
\newcommand{\tth}{\tilde{\theta}}
\newcommand{\cb}{{\cal B}}
\newcommand{\cg}{{\cal G}}

\title{Curvature correction to the field emission current}
 

\author{Debabrata Biswas}
\affiliation{
Bhabha Atomic Research Centre,
Mumbai 400 085, INDIA}
\affiliation{Homi Bhabha National Institute, Mumbai 400 094, INDIA}
\author{Rajasree Ramachandran}
\affiliation{
Bhabha Atomic Research Centre,
Mumbai 400 085, INDIA}


\begin{abstract}
 The curvature of field emitter tips leads to an altered tunneling potential that assumes
 significance when the radius of curvature is small. We provide here an analytical curvature-corrected
 formula for the field emission current from smooth vertically aligned emitter tips and test its applicability
 across a range of apex radius, $R_a$, and local electric field, $E_a$. It is found to give
 excellent results for $R_a > 10$nm with errors generally less than $10\%$.
 Surprisingly, for the uncorrected potential, we find the errors to be high even at $R_a = 100$~nm
 ($ > 35\%$ ~at $E_a = 3$~V/nm)  and
 conclude that curvature correction is essential for apex radius less than a micron.

\end{abstract}






\maketitle




\section{Introduction}
\label{sec:intro}

With the increasing use of nanostructured materials such as carbon nanotubes,
nanowires and nanocones \cite{teo,parmee,spindt91,lee2002,baylor}  in field emission cathodes,
the need for an extension to the Fowler-Nordheim (FN) formalism
\cite{FN,Nordheim,murphy,forbes,forbes_deane,jensen_ency} to deal with emission from nano-tipped emitters
is recognized \cite{KX,jensen_image,fursey,db_imag,db_ext}. 
For emitters with apex radius of curvature $R_a$ in the nanometer
regime, there are competing influences that determine the net field emission current
from an isolated emitter. While the local electric field on the emitter surface
increases with $1/R_a$ due to increased field enhancement, 
the tunneling potential on the other hand becomes wider as corrections to the external and image
potential start contributing. These factors influence the local current density on the
emitter surface, but, for purposes of determining the net emission current, the
decrease in net emission area with increasing curvature must also be factored in.
This interplay has added complexity when a large area field
emitter (LAFE) is studied \cite{db_rudra} since
shielding effects and the spatial density of emitters become important in determining
the net emitted current.

At the zeroth level, curvature effects are incorporated in the planar
FN formalism in terms of the apex field enhancement factor $\gamma_a$ such that the local
field  $E({\bf{r}}) = \gamma({\bf{r}}) E_0$ at a point ${\bf{r}}$ on the emitter surface. Here
$E_0$ is the macroscopic or asymptotic field far away from the
emitter tip. Thus, the current density is expressed as \cite{murphy,forbes_deane}

\be
J({\bf{r}}) =  \frac{1}{t_F^2} \frac{A_\fn}{W} E^2({\bf{r}}) 
e^{-\mathlarger{B}_\fn \mathlarger{\nu}_F W^{3/2}/E({\bf{r}})}. \label{eq:FN}
\ee

\noi
In the above, the free electron model is assumed and barrier
lowering due to the image potential is incorporated so that the net
potential experienced by the electron (Schottky-Nordheim barrier \cite{Schottky,Nordheim}) is

\be
V(s)  = \phi - e E s - \frac{e^2}{16 \pi\epsilon_0 s} \label{eq:pot0}
\ee

\noi
where $E$ is the local field on the surface of the emitter and $s$
measures the distance normal to the emitter surface.
In Eq.~\ref{eq:FN} above, $A_\fn~\simeq~1.541434~{\rm \mu A~eV~V}^{-2}$ and 
$B_\fn~\simeq~6.830890~{\rm eV}^{-3/2}{\rm V~nm}^{-1}$ are the conventional FN constants,
$\phi = \En_F + W$ where $W$ is the work function  and $\En_F$ the Fermi energy (both $W$ and $\En_F$ in eV), while \cite{forbes,forbes_deane}

\bea
\mathlarger{\nu}_F({\bf r}) = \mathlarger{\nu}(f_F({\bf r})) & \simeq & 1 - f_F + \frac{1}{6}f_F\ln f_F \\
\mathlarger{t}_F({\bf r})  = \mathlarger{t}(f_F({\bf r})) & \simeq & 1 + f_F/9 - \frac{1}{18}f_F\ln f_F
\eea

\noi
are correction factors due to the
image potential, with $f_F = f_F({\bf r}) \simeq 1.439965 E({\bf{r}})/W^2$.

Eq.~\ref{eq:FN} serves well to analyze experimental data under conditions compatible
with the model assumptions as well for curved emitters with radius of curvature
much larger than the tunneling distance \cite{lambda,Mayer}.
When the emitter tip radius is small, the flat-emitter assumption breaks down. It is not
very clear at what apex radius of curvature this occurs even though some  studies
suggest \cite{KX} that this could be around $R_a = 20$nm. Our investigations here reveal
that curvature dependent corrections are essential for radius of curvature as large as $R_a = 1~\mu$m
when the net emission current is of interest and the acceptable error is $5\%$.

The first few curvature correction terms to the tunneling potential of Eq.~\ref{eq:pot0} are now known for any point near
the emitter apex \cite{KX,fursey,jensen_image,db_imag,db_ext}. These can be incorporated to obtain a
suitably corrected expression for the tunneling transmission coefficient and
hence the current density. It is assumed here that the field lines 
can be considered linear in the tunneling region even for curved emitters
and 1-dimensional semiclassics continues to hold. 
For very small apex radius of curvature ($R_a < 5$~nm) however, the field lines are likely
to be curved even in the tunneling region and the curvature corrections in the potential must be
accompanied by a multi-dimensional tunneling treatment. There are other factors in a
real system that are not accounted for in this simplified picture. For instance, field electron
emission must be treated in conjunction with space-charge effects which can further distort
tunneling paths. We shall ignore these complications and merely assume that the field
lines are approximately linear and along the normal to the emitter surface in the tunneling regime.

In the following, we shall briefly review the corrections to the tunneling potential and deal with
the corresponding corrections to the transmission coefficient, the current density and
the net emitted current. This is followed by our numerical results for the hemi-ellipsoid
where exact results are known. In the rest of this paper, we shall consider an axially
symmetric vertically aligned emitter, parallel to the direction of the asymptotic electrostatic
field, $E_0 \hat{z}$.

\section{Curvature corrections}

\subsection{The tunneling potential}

The tunneling potential in Eq.~\ref{eq:pot0} is appropriate when the local radius of curvature
is very large and curvature terms may be neglected. In general, both
the image and external potential get modified. The image 
potential takes the form \cite{fursey,jensen_image,db_imag}

\be
- \frac{e^2}{16 \pi\epsilon_0 s} (1 - \frac{s}{2R_2} + \ldots) \simeq 
- \frac{e^2}{16 \pi\epsilon_0 s (1 + \frac{s}{2R_2})} \label{eq:potimag1}
\ee

\noi
where $R_2$ is the second principle radius of curvature at a point on the
emitter surface. Note that at the apex, $R_2 = R_a$ where $R_a$ is the apex radius
of curvature. Eq.~\ref{eq:potimag1} is essentially a local spherical
approximation of the emitter surface and reflects the interaction between
the electron at a distance $s$ from the emitter and its image charge
having magnitude $eR_2/(R_2+s)$ placed at a distance $R_2^2/(R_2 + s)$ from
the centre of the sphere.

The external potential also changes in the tunneling region when the
local radius of curvature is small. As mentioned
earlier however, we shall continue to treat the field lines as approximately
linear, even though we shall account for the change in magnitude of the field
in the tunneling region. This can be partially justified by noting that at
the apex of the vertically aligned emitter, the
field line continues to be along the axis (hence linear) even as the magnitude of the field drops
away from the apex. Since field emission in sharp emitters occurs predominantly
near the apex (small effective emission area), the linearity approximation is
justified.

The first order curvature corrected external potential at the apex \cite{KX}
for a general axially symmetric emitter is

\be
V_{ext}^{(1)}(s) =  - e E s \Big[1 - \frac{s}{R_a} \Big].
\ee

\noi
For points ${\bf r}$ close to the apex, 
analytical studies of the hemi-ellipsoid and hyperboloid emitters and numerical
evidence from other geometries (such as the conical emitter) show that
the external potential is well represented in the tunneling region  by \cite{db_ext}

\be
V_{ext}(s) =  - e E s \Big[1 - \frac{s}{R_2} + \frac{4}{3} \big(\frac{s}{R_2}\big)^2 \Big]
\ee

\noi
where $R_2$ is the second principle radius of curvature.
The net curvature corrected tunneling potential is thus

\be
V_C(s)  = \phi - e E s \Big[ 1 - \frac{s}{R_2} + \frac{4}{3} \big(\frac{s}{R_2}\big)^2 \Big] - \frac{e^2}{16 \pi\epsilon_0 s (1 + \frac{s}{2R_2})}. \label{eq:potC2}
\ee

\noi
It is implicit here that the potential depends on the position ${\bf r} = (\rho,z)$ on the
axially symmetric emitter
surface through the local field $E = E({\bf r})$ and the radius of curvature $R_2 = R_2({\bf r})$.
For vertically aligned emitters \cite{db_ultram}

\bea
E({\bf r}) &  = & E_a \cos\ttl,~~ \text{where} \label{eq:Evariation} \\
\cos\ttl & = & \frac{z/h}{\sqrt{(\rho/R_a)^2 + (z/h)^2}}
\eea

\noi
where $E_a$ is the local field at the apex and

\be
R_2 = R_a \Big[ 1 + \big(\frac{\rho}{R_a}\big)^2\Big]^{1/2}.
\ee

\noi
It is assumed that the tip is smooth and can be expressed locally as
$z = h - \rho^2/(2R_a)$. With these additional inputs, the net field emission
current can in principle be calculated for highly curved emitter tips.

In the following, we shall use the form of the tunneling potential
as given in Eq.~\ref{eq:potC2}. However, for the sake of comparison, we shall also
use the first order correction and denote it by

\be
V_C^{(1)}(s)  = \phi - e E s \Big[ 1 - \frac{s}{R_2}  \Big] - \frac{e^2}{16 \pi\epsilon_0 s (1 + \frac{s}{2R_2})}. \label{eq:potC1}
\ee

\subsection{The curvature corrected current density}

Assuming a free electron model, the current density is evaluated at zero temperature as

  \be
  J = \frac{2me}{(2\pi)^2 \hbar^3} \int_0^{\En_F} T({\cal E})(\En_F  - {\cal E}) d{\cal E}   \label{eq:Jbasic}
  \ee

\noi
where $T({\cal E})$ is the transmission coefficient at electron energy ${\cal E}$ measured with
respect to the bottom of the conduction band,
$m$ is the mass of the electron, $e$ is the
magnitude of the electron charge and $\En_F$ is the Fermi level. Eq.~\ref{eq:FN}
follows (i)  on using the WKB expression for transmission coefficient

\be
T_{WKB}({\cal E}) = \exp\Big(-\frac{2}{\hbar} \sqrt{2m} \int_{s_1}^{s_2} \sqrt{V(s) - \En}~ds \Big), \label{eq:TCwkb} 
\ee

\noi
in Eq.~\ref{eq:Jbasic} with $V(s)$ given by Eq.~\ref{eq:pot0}, (ii) approximating the integral above as

\be
\int_{s_1}^{s_2} \sqrt{V(s) - \En}~ds  \simeq
\frac{2}{3} \frac{(\phi - \En)^{3/2}}{eF} \mathlarger{\nu}(f)  \label{eq:GamowG}
\ee

\noi
with $f  \simeq   1.439965 E({\bf{r}})/(\phi - \En)^2$ and finally (iii) Taylor expanding
it about the Fermi energy in order to carry out the energy integration. In the above
$s_1,s_2$ are the roots of $V(s) - \En = 0$.

For the curvature corrected tunneling potential $V_C(s)$, a similar procedure
can be followed. Following Ref.~[12],
the curvature corrected current density for the potential $V_C$
can be expressed  by replacing $\nu_F$ and $t_F$ respectively
in Eq.~\ref{eq:FN} by $\tilde{\nu}_F$ and $\tilde{t}_F$ \cite{apex_same}.  The curvature corrected current density at any point
around the apex is thus \cite{first_second_same}

\be
J_C({\bf{r}}) =  \frac{1}{(\mathlarger{\tilde{t}}_F)^2} \frac{A_\fn}{W} E^2({\bf{r}}) 
e^{-\mathlarger{B}_\fn \mathlarger{\tilde{\nu}}_F W^{3/2}/E({\bf{r}})}. \label{eq:FNC}
\ee

\noi
where

\bea
\mathlarger{\tilde{\nu}}_F({\bf r}) & = & \mathlarger{\nu}_F({\bf r})  +   \frac{W}{E({\bf r}) R_2({\bf r})} w_F({\bf r})~~ \\
\mathlarger{\tilde{t}}_F({\bf r}) & = & \mathlarger{t}_F({\bf r})  +   \frac{W}{E({\bf r}) R_2({\bf r})}  \psi_F({\bf r})~~ \\
f_F & = & f_F({\bf r}) \simeq 1.439965 \frac{E({\bf{r}})}{W^2}  
\eea

\noi
and

\bea
w_F({\bf r}) = w(f_F({\bf r})) & = & \frac{4}{5} - \frac{7}{40}f_F - \frac{1}{200} f_F \ln(f_F) \\
\psi_F({\bf r}) = \psi(f_F({\bf r})) & = & \frac{4}{3} - \frac{1}{500} f_F - \frac{1}{30}f_F \ln(f_F).
\eea

The adequateness of Eq.~\ref{eq:FNC} as a curvature-corrected current density is best tested
on evaluation of the net emitted current. We shall therefore postpone a discussion on its merits
till the next section.

\subsection{The curvature corrected emission current}

The current from an emitter tip can be evaluated by integrating the current density over the
emitter surface:

\be
I = \int J({\bf{r}}) 2\pi \rho \sqrt{1 + (dz/d\rho)^2} d\rho
\ee

\noi
where ${\bf{r}} = (\rho,z)$. For smooth axially symmetric vertically
aligned emitters, $z = h - \rho^2/(2R_a)$ near the tip. Thus,

\be
I =  \int J({\rho}) 2\pi \rho \sqrt{1 + (\rho/R_a)^2} d\rho \label{eq:Irho}
\ee

\noi
or, alternately, in terms of the normalized angle\cite{db_distrib}  $\ttl$ , 

\be
I  \simeq  2\pi R_a^2 \int J(\ttl)~\frac{\sin\ttl}{\cos^4\ttl} d\ttl.  \label{eq:Itheta}
\ee

\noi
Numerical evaluation of the current $I$ can be performed by typically integrating Eq.~\ref{eq:Irho}
from 0 to $R_a$ using appropriate forms of the current density.

We shall first dwell on the necessity and domain of applicability of the corrections 
to the tunneling potential. In order to establish this numerically, consider a hemi-ellipsoid
on a grounded conducting plane in the presence of an asymptotic field $E_0 \hat{z}$.
The exact analytical form of the potential for this system is well known.
The various forms of the tunneling potential that we shall compare with are (i) zeroth order
as in Eq.~\ref{eq:pot0} (ii) first order as in Eq.~\ref{eq:potC1} and
(iii) second order as given in Eq.~\ref{eq:potC2}. The errors in net emission current
for these potentials can be computed relative to the analytical tunneling potential
along (a) the normal to the surface (b) along the field line \cite{approx_fieldline}. As remarked earlier,
the normal to the surface and the field line approximately coincide in the tunneling
region if the curvature is not too sharp.

\begin{figure}[thb]
\vskip -1.75 cm
\hspace*{-1.0cm}\includegraphics[width=0.6\textwidth]{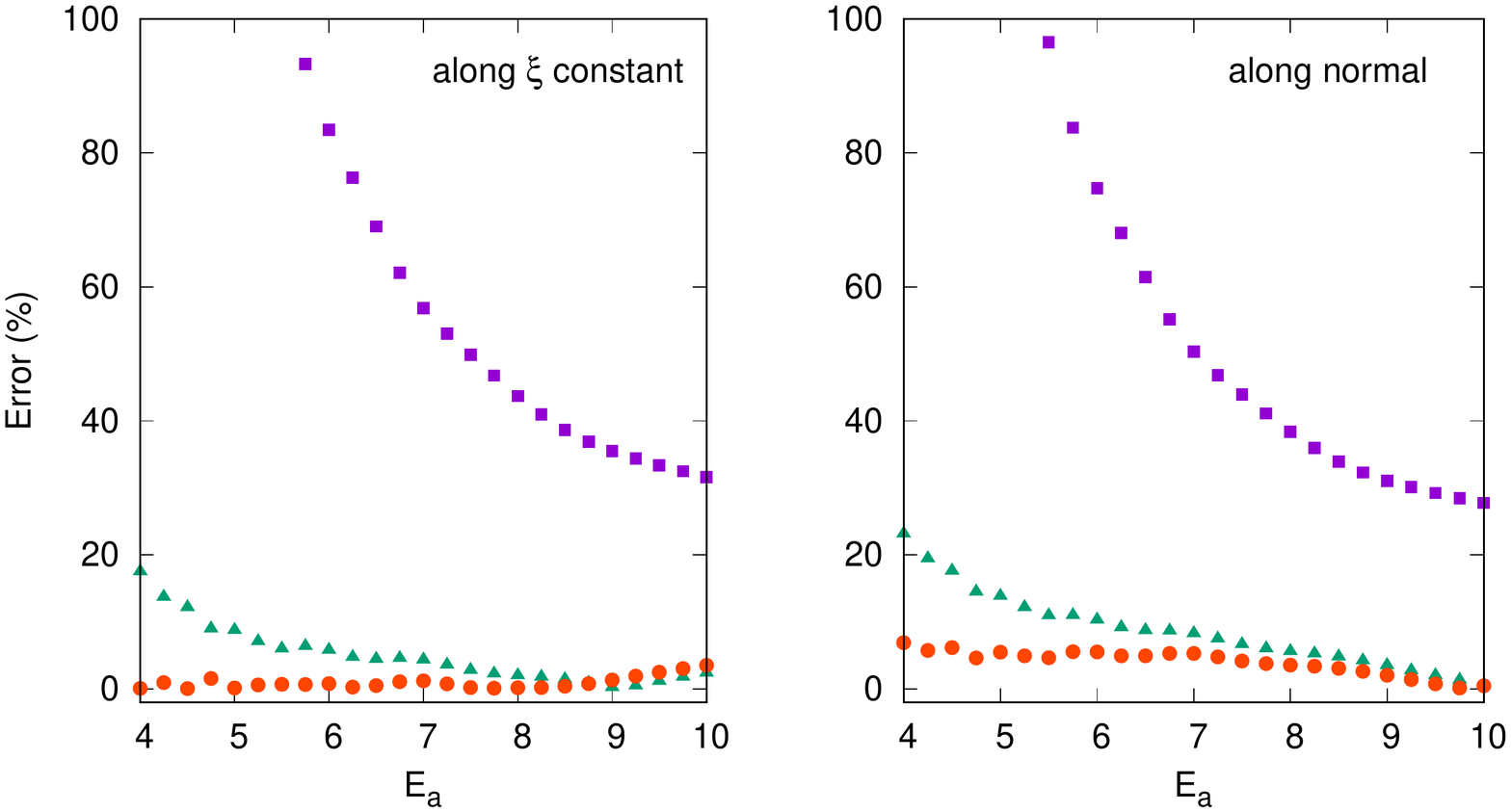}
\vskip -0.9 cm
\caption{The error in net emission current  using the
  $0^{th}~ \text{(squares)}$, $1^{st}~ \text{(triangle)}$
 and $2^{nd}~ \text{(circle)}$ order tunneling
  potentials relative to the analytical tunneling potential computed along
  field lines ($\xi = \text{constant}$,left) and along normal to the surface (right). The apex
radius of the hemiellipsoid $R_a = 13$~nm.}
\label{fig:error_pot012}
\end{figure}

Fig.~\ref{fig:error_pot012} shows the error in net emission current using the zeroth order
(Eq.~\ref{eq:pot0}, filled squares), first order (Eq.~\ref{eq:potC1}, filled triangle)
and second order (Eq.~\ref{eq:potC2},filled circle)) tunneling potentials. 
The errors are computed relative to the
current found using the analytical potential (a) along field lines ($\xi = \text{constant}$ where
($\eta,\xi,\varphi$) are prolate spheroidal co-ordinates \cite{approx_fieldline,smythe} )
and (b) along normal. In all cases, the transfer matrix method \cite{DBVishal} is adopted for the
transmission coefficients. Note that at $R_a = 13$~nm, the error is large for the
first order tunneling potential at lower local apex field strengths. For smaller apex
radius, the errors for both the zeroth and first order tunneling potential are much larger.

\begin{figure}[thb]
\vskip -1.75 cm
\hspace*{-1.0cm}\includegraphics[width=0.6\textwidth]{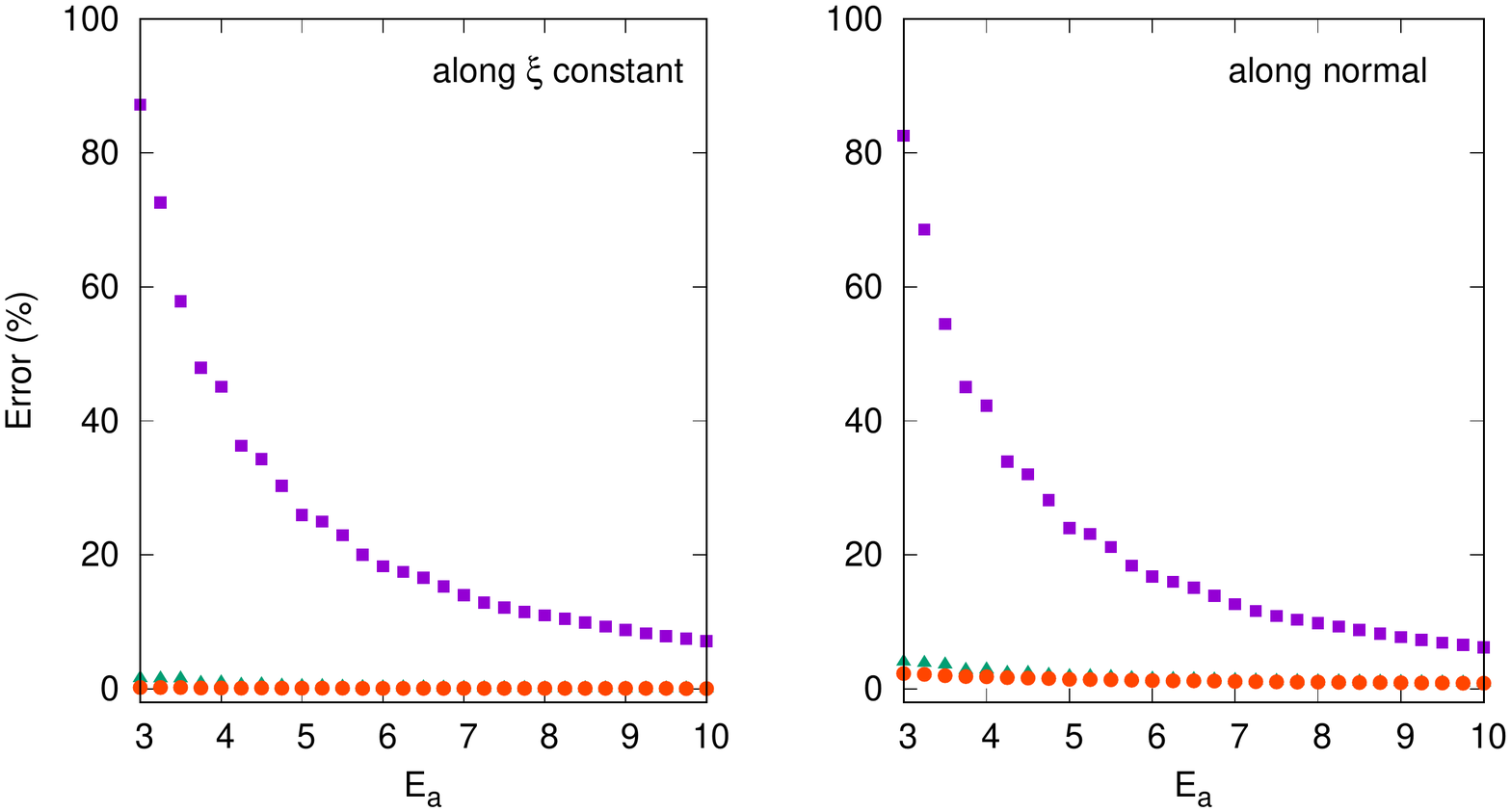}
\vskip -0.9 cm
\caption{As in Fig.~\ref{fig:error_pot012} for apex
radius of curvature $R_a = 50$~nm.}
\label{fig:error_pot012_50}
\end{figure}

\begin{figure}[thb]
\vskip -1.75 cm
\hspace*{-1.0cm}\includegraphics[width=0.6\textwidth]{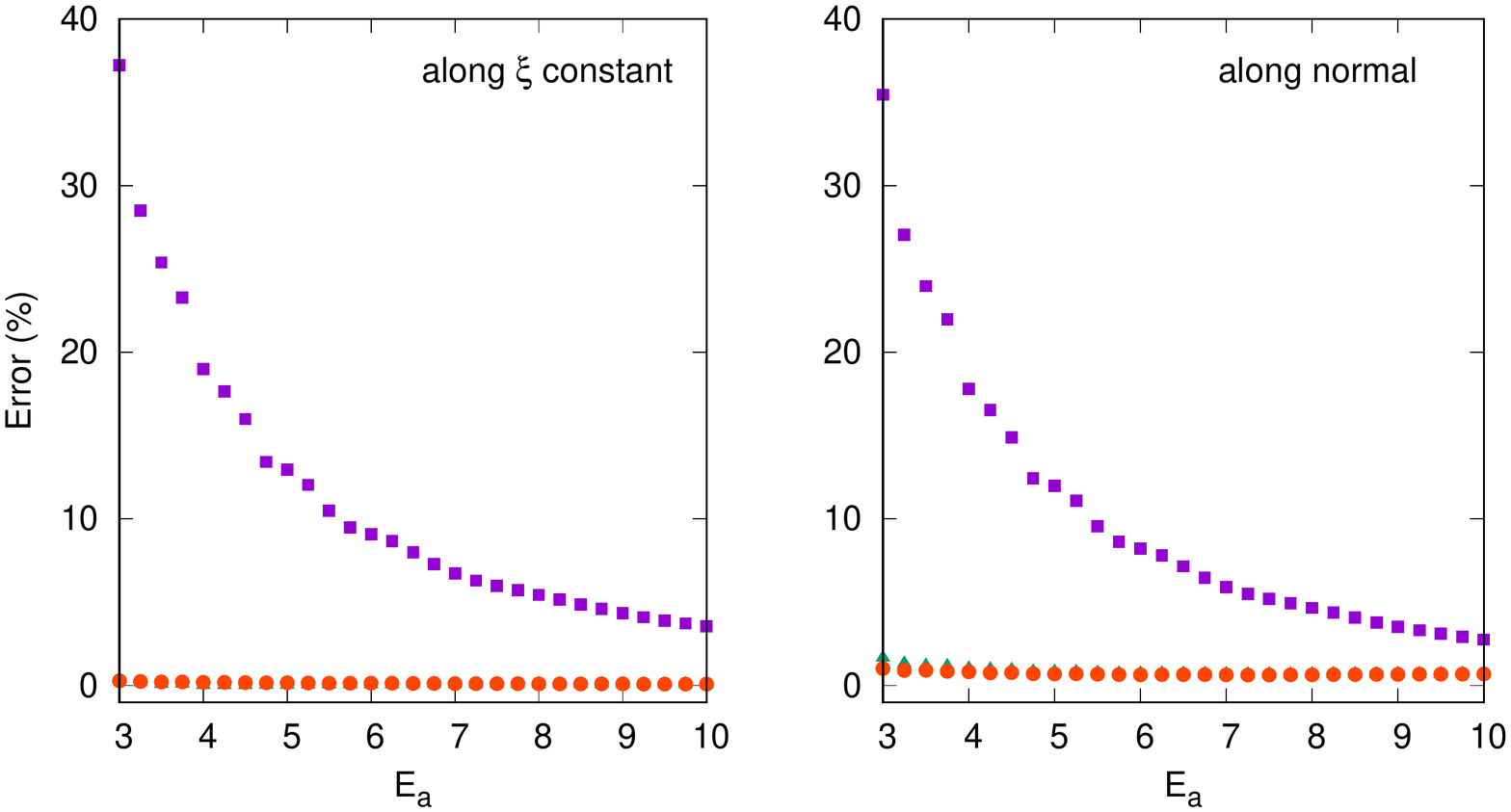}
\vskip -0.9 cm
\caption{As in Fig.~\ref{fig:error_pot012} for $R_a = 100$~nm.}
\label{fig:error_pot012_100}
\end{figure}

At $R_a = 50$~nm, it is still worthwhile to use the second order correction to the
tunneling potential for lower apex fields as seen in Fig.~\ref{fig:error_pot012_50}. The error
for the zeroth order decreases but remains large. At $R_a = 100$~nm (Fig.~\ref{fig:error_pot012_100}),
the first order and second order corrections are hard to distinguish even at lower
field strengths and while the error for the
zeroth order potential decreases further, the curvature-corrected potential must
still be used especially at lower fields. At $R_a = 500$nm, the error in using
the zeroth order potential falls to $7\%$ at $E_a = 3$~V/nm while at $R_a = 1~\mu$ m and $E_a = 3$~V/nm,
the error is 5.3\%. Thus, it is profitable to
use the second order corrected potential for $R_a < 1~\mu$m, especially at
lower field strengths.

\begin{figure}[tbh]
\vskip -1.0 cm
\hspace*{-1.0cm}\includegraphics[width=0.6\textwidth]{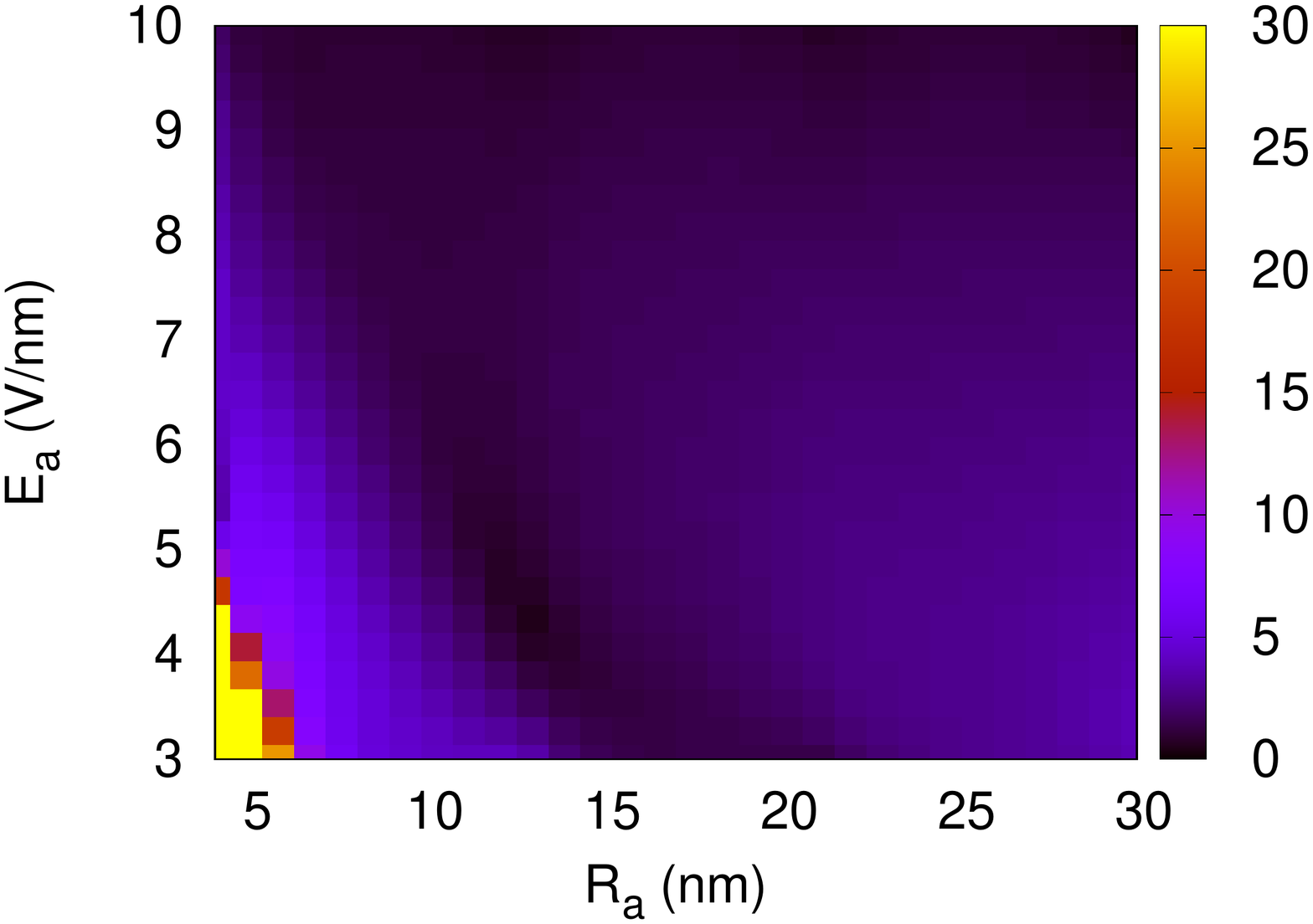}
\vskip -1.0 cm
\caption{The relative \% error in the net emission current calculated using  Eq.~\ref{eq:FNC}
  for ellipsoidal emitters with varying apex radius of curvature $R_a$ and local apex field $E_a$. The error
  is relative to the option (c) using Eq.~\ref{eq:Jwkbexpand}.}
\label{fig:error_krysTC_wkbexpand}
\end{figure}

Having established that the second order tunneling potential (Eq.~\ref{eq:potC2})
is essential for curved emitters when $R_a < 1~\mu$m,  
we next turn our attention to the efficacy of the curvature corrected current
density of Eq.~\ref{eq:FNC}. The total emitted current can be calculated using
Eq.~\ref{eq:Irho} along with Eq.~\ref{eq:FNC} and the value obtained can be compared using one
of the following alternative (and more exact) methods of obtaining the current density:
(a) transfer matrix or equivalent ``exact''
evaluation of the transmission coefficient \cite{DBVishal} at all energies and exact energy integration (b)
an exact WKB evaluation of the transmission coefficient (Eq.~\ref{eq:TCwkb} with $V$ replaced
by $V_C$) at all energies and
an exact energy integration (c) Taylor expansion (upto the linear term)
of Eq.~\ref{eq:TCwkb} around the Fermi energy and exact evaluation of all the integrals.
Clearly, the current evaluated using Eq.~\ref{eq:FNC} is expected to be closest to
option (c) above. The current density in option (c) can be expressed as

\begin{figure}[hbt]
\vskip -0.75 cm
\hspace*{-1.0cm}\includegraphics[width=0.6\textwidth]{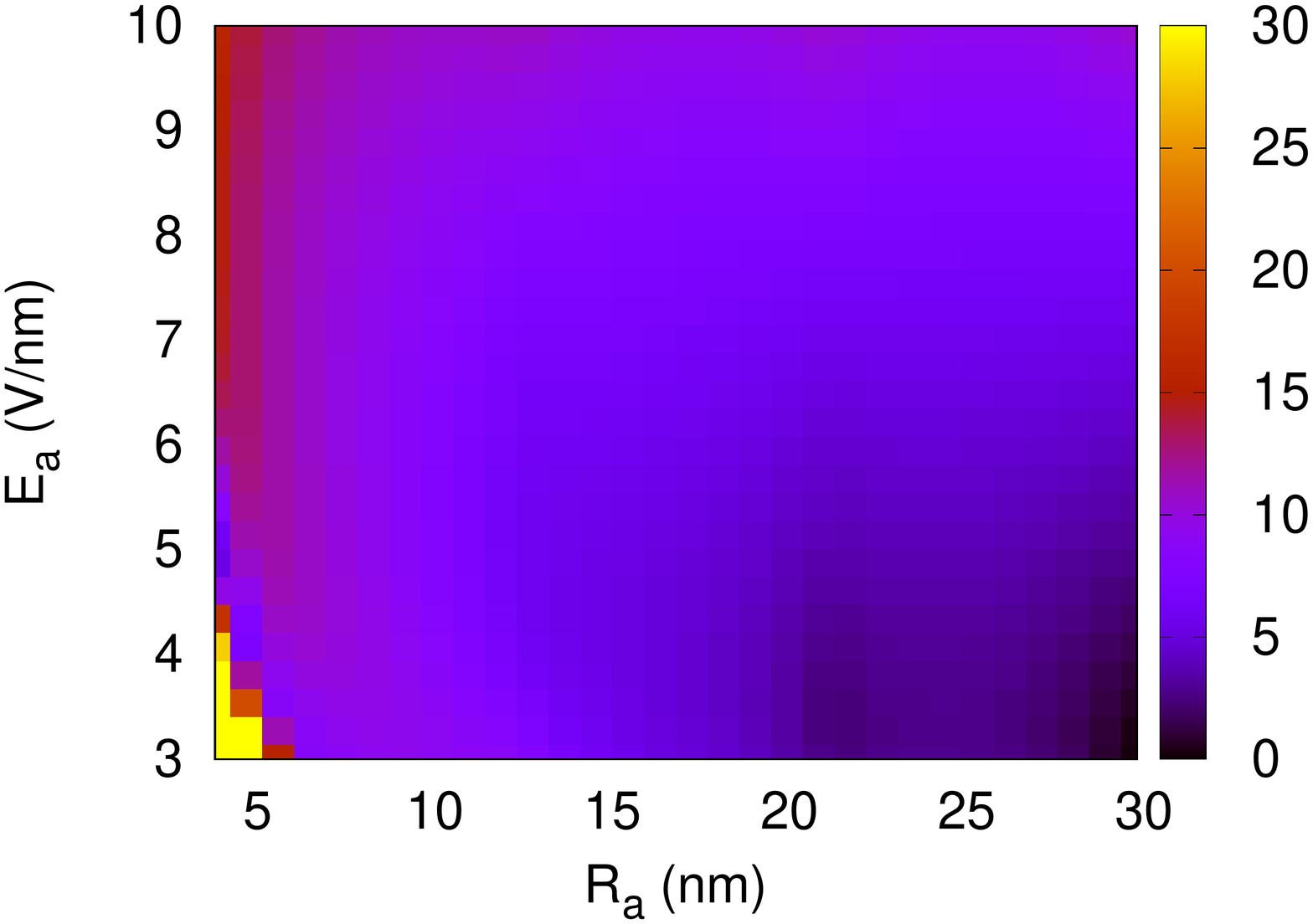}
\vskip -0.9 cm
\caption{The relative \% error in the net emission current calculated using Eq.~\ref{eq:FNC} relative to
  the  current determined using option (b) where the exact WKB transmission coefficient is used
  together with exact energy integration.}
\label{fig:error_krysTC_wkbexact}
\end{figure}

\be
\begin{split}
J = &\frac{2me}{(2\pi)^2 \hbar^3}  \int_0^{\En_F} (\En_F  - {\cal E}) \exp\Big[-g\int_{s_1}^{s_2} \sqrt{V_C(s) - E_F}\Big] \times \\ 
& \exp\Big[(E - E_F) \frac{g}{2} \int_{s_1}^{s_2} \frac{1}{\sqrt{V_C(s) - E_F}}~ds\Big]
 d{\cal E}   \label{eq:Jwkbexpand}
\end{split}
\ee

\noi
with the integrals evaluated numerically. Eq.~\ref{eq:Jwkbexpand} together with Eq.~\ref{eq:Irho}
gives the total emitted current. Here $s_1$ and $s_2$ are zeroes of $V_C(s) - E_F = 0$.

Fig.~\ref{fig:error_krysTC_wkbexpand} shows the error in net
emission current evaluated using $J_C$ (Eq.~\ref{eq:FNC}) relative to that obtained
using Eq.~\ref{eq:Jwkbexpand}. 
Clearly, the relative error is small for $R_a > 7$nm at all field strengths considered while for smaller
apex radius, error is larger when $R_a E_a$ is small as expected from the nature of the correction.

We next compute the error in net emission current evaluated using $J_C$ relative to option (b)
above where the exact WKB transmission coefficient is used together with the exact energy
integration. Note that  the  exact WKB transmission coefficient is determined by
evaluating the Gamow exponent 

\be
G = g \int_{s_1}^{s_2} \Big(V_C(s) - \En \Big)^{1/2}
\ee

\noi
exactly.  Here $g = 2\sqrt{2m}/\hbar \simeq 10.246~(\text{eV})^{-1/2} (\text{nm})^{-1}$. The integral
is performed numerically in order to determine $G$ and hence the transmission coefficient.
The relative error is shown in Fig.~\ref{fig:error_krysTC_wkbexact}. The error
is somewhat larger compared to the previous case since the energy integration
is exact in option (b) while Eq.~\ref{eq:FNC} uses a Taylor expansion
in energy. For $R_a > 10$nm however, the error is reasonably small. 
Note also that option (b) is close to the exact transfer matrix result calculated using
option (a) with errors within $1\%$ in the region of interest.

\subsection{An approximate analytical expression for net current}

For an approximate analytical expression, Eq.~\ref{eq:Itheta} can be used together with
the curvature corrected current density $J_C$ expressed in terms of $\ttl$ as \cite{db_distrib}

\be
J(\ttl) =  \frac{1}{\tilde{\mathlarger{t}}_F(\ttl)^2} \frac{A_\fn}{W} E_a^2 \cos^2\ttl~
e^{-B_\fn \tilde{\mathlarger{\nu}}_F(\ttl) W^{3/2}/(E_a \cos\ttl)}. \label{eq:FNC1}
\ee

\noi
Eq.~\ref{eq:FNC1} follows on using Eq.~\ref{eq:Evariation} in Eq.~\ref{eq:FNC}. Thus,

\be
I  \simeq  \cc \int_0^{\pi/3} \frac{\sin\tth}{\cos^2\tth} \frac{1}{\mathlarger{\tilde{t}}_F^2(\tth)} e^{-\cb \mathlarger{\tilde{\nu}}_F(\tth)/\cos\tth} d\tth  \label{eq:IFNC}
\ee

\noi
where

\bea
\cc & = &   2\pi R_a^2 \frac{A_\fn}{W} E_a^2~~~\text{and} \\
\cb & = & \frac{B_\fn W^{3/2}}{E_a}
\eea

\noi
With the substitution $1/\cos\tth = 1 + x$, Eq.~\ref{eq:IFNC} reduces to \cite{db_distrib}

\be
I \simeq \cc \int_0^1 \frac{1}{{\mathlarger{\tilde{t}}_F^2(x)}} e^{-\cb \mathlarger{\tilde{\nu}}_F(x) (1 + x)} dx. \label{eq:totI1} 
\ee

\begin{figure}[htb]
\vskip -0.5 cm
\hspace*{-1.0cm}\includegraphics[width=0.6\textwidth]{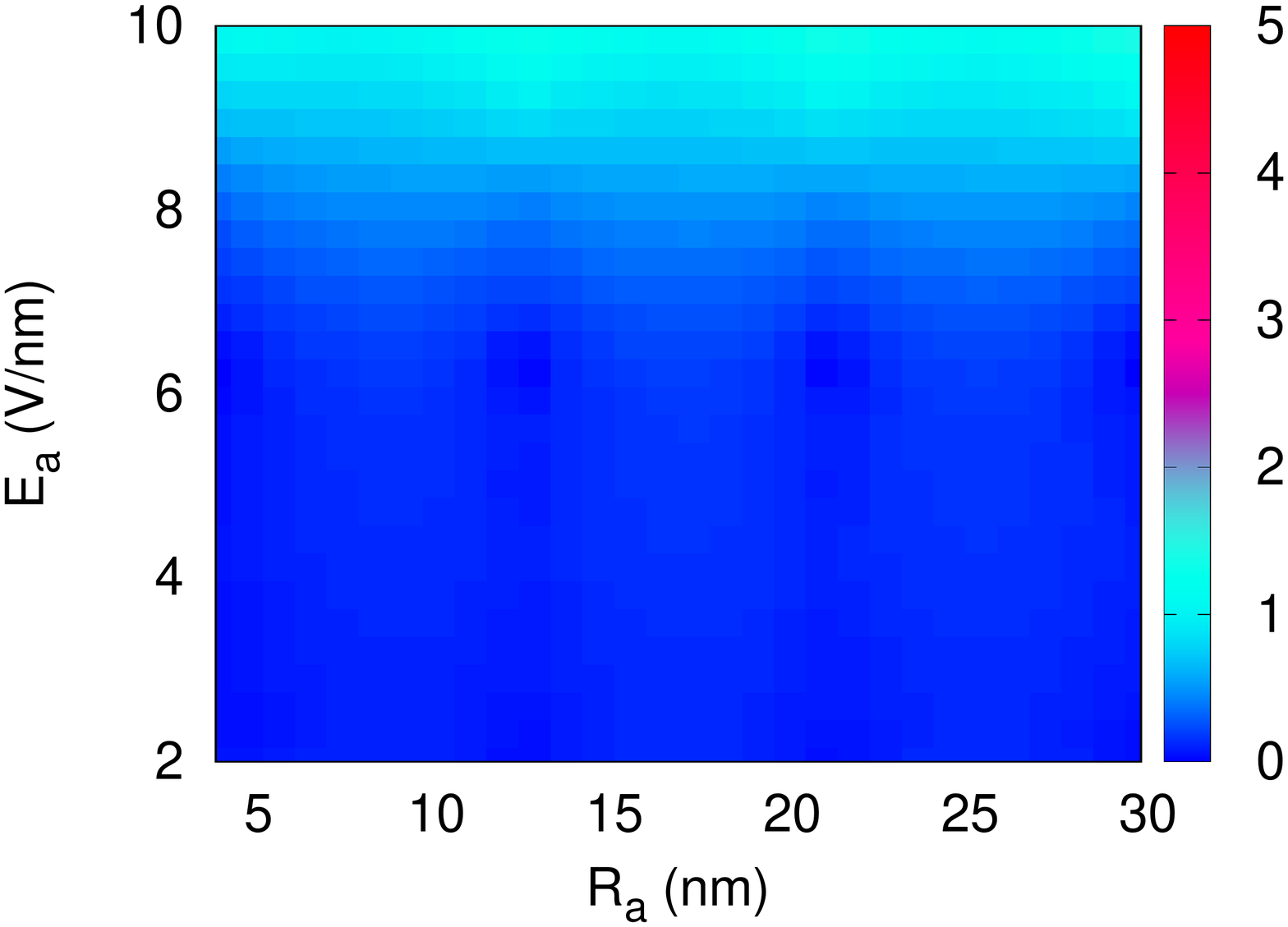}
\vskip -0.9 cm
\caption{The relative \% error in the net emission current calculated
  using Eqns.~\ref{eq:Iana} and \ref{eq:gfact0},
relative to emission current calculated using Eq.~\ref{eq:FNC}.}
\label{fig:gfact_Jkrys}
\end{figure}

\noi
Since the dominant contribution comes from the neighbourhood of $x = 0$, a Taylor expansion
of  $\mathlarger{\tilde{\nu}}_F(x) (1 + x)$ and $1/\mathlarger{\tilde{t}}_F^2(x)$ at $x = 0$
can be used. Keeping only the linear term\cite{db_distrib},

\bea
\mathlarger{\tilde{\nu}}_F(x) (1 + x) & = & \tilde{D}_0 + \tilde{D}_1 x + \mathcal{O}(x^2) \\
\frac{1}{\mathlarger{\tilde{t}}_F^2(x)} & = & \tilde{F}_0 + \tilde{F}_1 x +  \mathcal{O}(x^2)
\eea

\noi
where

\bea
\tilde{D}_0 &  = & \mathlarger{\nu}_F(0) +  {\cal X}_{F} w_F(0)  \\
\tilde{D}_1 & = & \big(1 - \frac{f_0}{6}\big) + {\cal X}_{F} \big(\frac{4}{5} + \frac{f_0}{200}\big) \\
\tilde{F}_0 & = & \frac{1}{\big[t_F(0) + {\cal X}_{F} \psi_F(0) \big]^2}  \\
\tilde{F}_1 & = &  \frac{ f_0 - f_0 \ln f_0 + {\cal X}_{F} \frac{3}{5} \big(\frac{53}{50}f_0 + f_0 \ln f_0 \Big) }{9 \big[t_F(0) + {\cal X}_{F} \psi_F(0) \Big]^3}
\eea

\noi
with $ {\cal X}_{F} = W/(E_a R_a)$ and $f_0 = 1.439965 E_a/W^2$. Note that the quantities, $\mathlarger{\nu}_F(0)$, $w_F(0)$,
$t_F(0)$ and $\psi_F(0)$ are calculated at the apex. Thus,

\be
I \simeq 2\pi R_a^2 J_C(0) \tilde{\cg}   \label{eq:Iana}
\ee

\noi
where

\be
\tilde{\cg}  = \frac{1}{\cb \tilde{D}_1} + \frac{\tilde{F}_1}{\tilde{F}_0} \frac{1}{(\cb \tilde{D}_1)^2}
\ee

\noi
determines the effective emission area at the apex. In effect, the first term 

\be
\tilde{\cg} \simeq \frac{1}{\cb \tilde{D}_1}  \label{eq:gfact0}
\ee

\noi
gives excellent results relative to the current derived using Eq.~\ref{eq:FNC}. A comparison
of the errors in current determined using Eqns.~\ref{eq:gfact0} and Eq.~\ref{eq:Iana} relative to the current
evaluated using Eq.~\ref{eq:FNC} is shown in Fig.~\ref{fig:gfact_Jkrys}. Clearly, the approximate
analytical formula does not introduce significant errors  as compared to a direct use of Eq.~\ref{eq:FNC}
for finding the current.

\begin{figure}[htb]
\vskip -0.5 cm
\hspace*{-1.0cm}\includegraphics[width=0.6\textwidth]{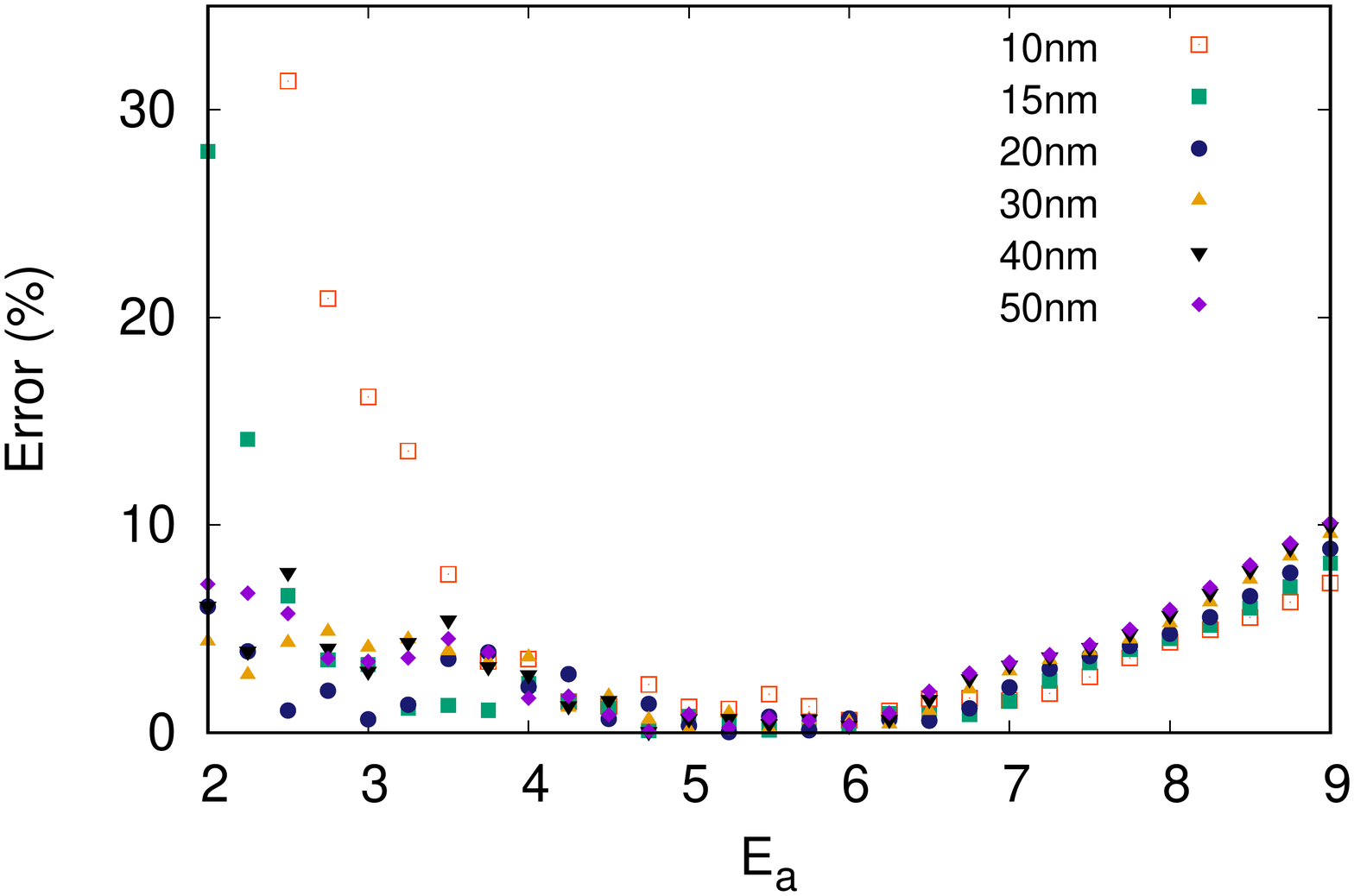}
\vskip -1.2 cm
\caption{Error in the emission current computed using Eqns.~\ref{eq:gfact0} and Eq.~\ref{eq:Iana}
  relative to the exact potential along the normal for various apex radius $R_a$. The local apex
field $E_a$ is measured in V/nm.}
\label{fig:gfact_normalTM}
\end{figure}

When compared to the current computed using the exact analytical potential along the normal,
the error is found to be generally below $10\%$ over a wide range of local apex field strengths
when $R_a > 10$nm as seen in Fig.~\ref{fig:gfact_normalTM}. The increase in
error away from $E_a \simeq 5$V/nm is likely due to the Taylor expansion in energy
as discussed earlier.

\begin{figure}[htb]
\vskip -0.5 cm
\hspace*{-1.0cm}\includegraphics[width=0.6\textwidth]{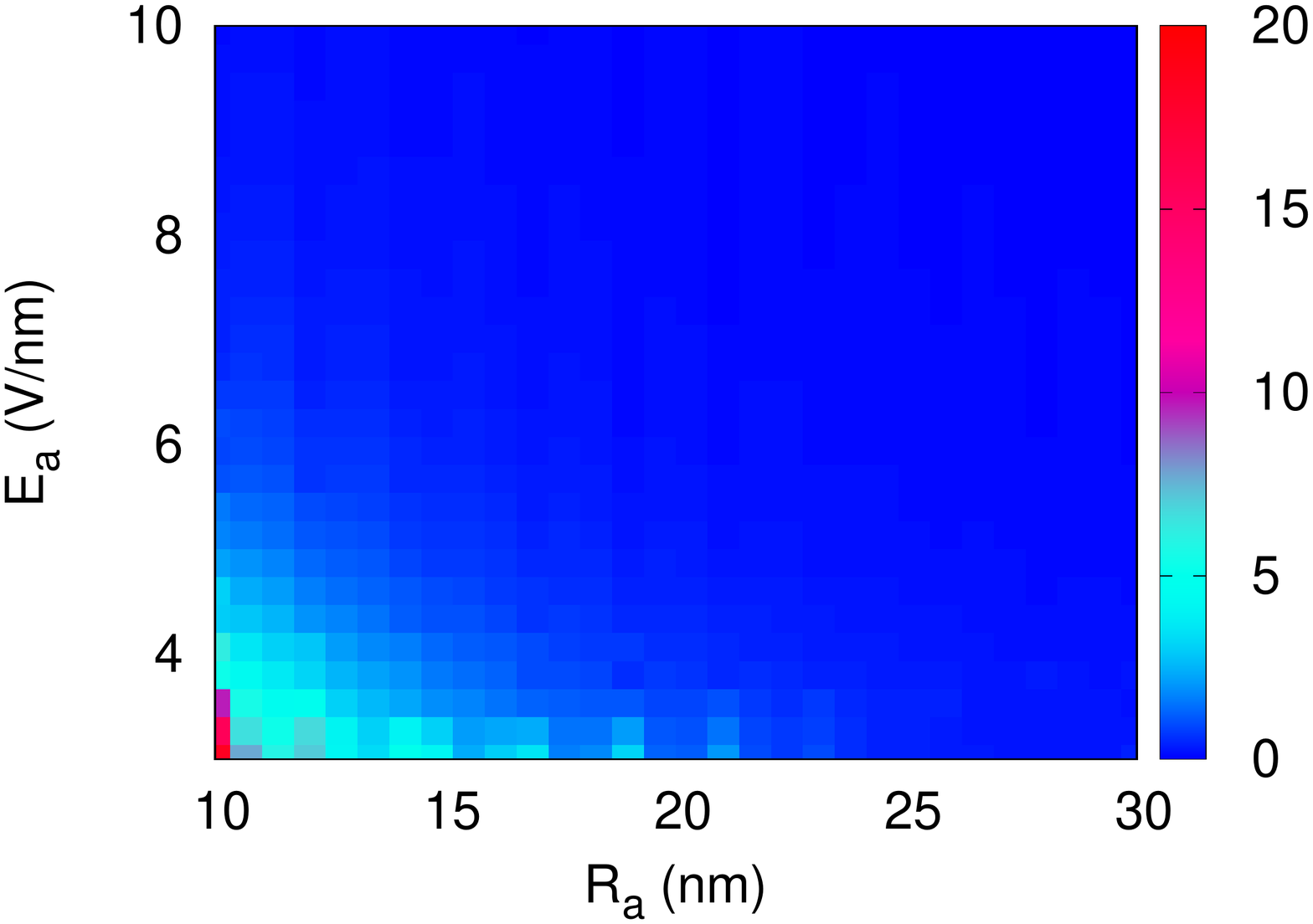}
\vskip -1.2 cm
\caption{Error in the emission current computed using the second order tunneling  potential
of Eq.~\ref{eq:potC2} relative to the exact potential along field lines.}
\label{fig:4b3_vs_xi}
\end{figure}

Finally, we revisit the domain of applicability of the second order potential $V_C(s)$.
A comparison of the currents obtained by transfer matrix
method using the exact potential along field lines ($\xi = \text{constant}$) and the second
order tunneling potential of Eq.~\ref{eq:potC2} is shown as an error map in Fig.~\ref{fig:4b3_vs_xi}.
The errors increase sharply for $R_a$ less than $10$~nm especially for the smaller apex fields
and is therefore not shown. The errors for $R_a > 10$~nm are generally below $5\%$
thus validating the use of the second order curvature corrected potential.

\section{Conclusions}

We have investigated the curvature corrected field emission current density and the net emission current
using a form of the tunneling potential that has been found to hold in analytically
solvable emitter models and numerically verified for other emitter shapes.
It was found that curvature correction is essential for apex radius of curvature smaller
than a micron with errors using the uncorrected potential as high as 37\% at $R_a = 100$~nm  and
around $5\%$ at $R_a = 1~\mu$m at $E_a = 3$~V/nm.

Using recent results, the curvature corrected current density at any point of the emitter surface was derived and
subsequently used to calculate the net emission current. This was used as a
basis for finding a simple analytical expression for the net emitted current from axially
symmetric vertically aligned smooth (parabolic) emitter tips. It was found that the
curvature corrected easy-to-use formula for emission current agreed to within 10\% of the
exact result for $R_a > 10$nm and to within $5\%$ if the transfer matrix method is used with
the curvature corrected potential.

\section{Acknowledgement}

The authors acknowledge several useful discussions with Dr. Raghwendra Kumar and Gaurav Singh.

\vskip -0.25 in
$\;$\\
\section{References} 



\begin{thebibliography}{99}

\bibitem{teo}  K.~B.~K.~Teo, E.~Minoux, L.~Hudanski, F.~Peauger, J.~P.~Schnell, L.~Gangloff, P.~Legagneux, D.~Dieumegard, G.~.A.J.~Amaratunga and W.~I.~Milne, Nature 437, 968 (2005).
\bibitem{parmee} R.~J.~Parmee, C.~M.~Collins, W.~I.~Milne, and M.~T.~Cole, Nano Convergence 2, 1 (2015).
\bibitem{spindt91} C.~A.~Spindt, C.~E.~Holland, A.~Rosengreen and I.~Brodie, IEEE Trans. on Electron Devices, 38, 2355 (1991).
\bibitem{lee2002} C.~J.~Lee et al., Appl. Phys. Lett. 81, 3648 (2002).
\bibitem{baylor} L.~R.~Baylor, V.~I.~Merkulov, E.~D.~Ellis, M.~A.~Guillorn, D.~H.~Lowndes, A.~V.~Melechko, M.~L.~Simpson, J.~H.~Whealton, J. Appl. Phys., 91, 4602 (2002).
\bibitem{FN} R.~H.~Fowler and L.~Nordheim, Proc. R. Soc. A 119, 173 (1928).
\bibitem{Nordheim} L.~Nordheim, Proc. R. Soc. A 121, 626 (1928).
\bibitem{murphy} E.~L.~Murphy and R.~H.~Good, Phys. Rev. 102, 1464 (1956).
\bibitem{forbes} R.~G.~Forbes, App. Phys. Lett. 89, 113122 (2006). 
\bibitem{forbes_deane} R.~G.~Forbes and J.~H.~B.~Deane, Proc. Roy. Soc. A 463, 2907 (2007).
\bibitem{jensen_ency} K.~L.~Jensen, {\it Field emission - fundamental theory to usage}, Wiley Encycl. Electr. Electron. Eng. (2014).
\bibitem{KX} A.~Kyritsakis and J.~P.~Xanthakis, Proc.~R.~Soc.~London, A471, 20140811 (2015).
\bibitem{jensen_image} K.~L.~Jensen, D.~A.~Shiffler, J.~R.~Harris, I.~M.~Rittersdorf, and J.~J.~Pettilo, J.~Vac.~Sci.~Technol., B 35, 02C101 (2017).
\bibitem{fursey} G.~N.~Fursey and D.~V.~Glazanov, J. Vac. Sci. Technol. B 16, 910 (1998).  
\bibitem{db_imag} D.~Biswas and R.~Rajasree, Phys. Plasmas, 24, 073107 (2017); 24, 079901 (2017).
\bibitem{db_ext} D.~Biswas, R.~Rajasree and G.~Singh, Phys. Plasmas 25, 013113 (2018).
\bibitem{db_rudra} D.~Biswas and R.~Rudra, {\it Shielding effects in random large area field emitters, the field enhancement factor distribution and current calculation} Phys. Plasmas (in press).
\bibitem{Schottky} W.~Schottky, Physik.~Zeitschr. 15,~872~(1914).
\bibitem{lambda} Even in this regime, Eq.~\ref{eq:FN} does not acurately predict the current density for all
  local fields, work function or Fermi energy, even though, the semiclassical transmission coefficient on
  which it is based, is reasonably accurate. This is due to a Taylor approximation
  of the transmission coefficient around the Fermi energy while performing the integration
  over all electron energies. More accurate results can however be computed numerically or using the table of coefficients
  in Mayer \cite{Mayer}  to account for a factor to be used alongside Eq.~\ref{eq:FN}.
\bibitem{Mayer} A.~Mayer, J.~Vac.~Sci.~Tech.~B, 29, 021803 (2011).
\bibitem{db_ultram} D.~Biswas, G.~Singh, S.~G.~Sarkar and R.~Kumar, Ultramicroscopy 185, 1 (2018).
\bibitem{db_distrib} D.~Biswas, Phys. Plasmas, 25, 043105 (2018).
\bibitem{apex_same} This is identical
  to the corrected apex current density of Ref.~[12] since $R_2 = R_a$ at the apex.
\bibitem{first_second_same} Since this in a Taylor expansion in $\frac{W}{E R_2}$
  evaluated at $R_2 \rightarrow \infty$, the coefficient of $\frac{W}{E R_2}$ is
  identical for $V_C$ and $V_C^{(1)}$.
\bibitem{approx_fieldline} Close to the vertically aligned hemi-ellipsoid surface, $\xi = {\text{constant}}$
  can be considered to be along the field line. Further away, the field lines deviate
  away from $\xi = {\text{constant}}$ and asymtotically point in the vertical direction.
\bibitem{smythe} W.~R.~Smythe, {\it Static and Dynamic Electricity}, Taylor and Francis (1989).
\bibitem{DBVishal} D.~Biswas and V.~Kumar, Phys.~Rev.~E 90, 013301 (2014).
\end{thebibliography}
\end{document}